\documentclass[final,5p,times,onecolumn]{elsarticle}

\usepackage{graphicx}
\usepackage{dcolumn}
\usepackage{bm}

\usepackage{amsmath}
\usepackage{amssymb}
\usepackage{mathptmx}
\usepackage{textcomp}
\usepackage{tabularx} 
\usepackage{times}
\usepackage{wrapfig}
\usepackage{caption}
\usepackage{subcaption}
\usepackage{adjustbox}
\usepackage{listings}
\usepackage{silence}
\usepackage{wasysym}
\usepackage{titlesec}
\usepackage{graphicx}
\usepackage{colortbl}
\usepackage[table,xcdraw]{xcolor}
\usepackage{booktabs}
\usepackage{chemfig}
\usepackage{graphicx}
\usepackage[utf8]{inputenc}
\usepackage[T1]{fontenc}
\usepackage{etoolbox}
\usepackage{csquotes}
\usepackage{xcolor}

\usepackage[table]{xcolor} 
\usepackage{hyperref}
\hypersetup{
    colorlinks=true,
    linkcolor=cyan,
    filecolor=magenta,      
    urlcolor=blue}

\definecolor{headergray}{rgb}{0.5, 0.5, 0.5}  
\usepackage{hyperref}

\begin{document}

\title{Effects of Galactic Irradiation on Thermal and Electronic Transport in Tungsten}

\author[LANL]{C.Ugwumadu}
\ead{cugwumadu@lanl.gov}
\affiliation[LANL]{organization = Physics of Condensed Matter and Complex Systems (T-4) Group, addressline = { Los Alamos National Laboratory}, city = { Los Alamos}, postcode = {87545}, state ={NM}, country={USA}}

\author[OU]{D. A. Drabold}
\ead{drabold@ohio.edu}
\affiliation[OU]{organization={Department of Physics and Astronomy, Nanoscale and Quantum Phenomena Institute (NQPI)},
            addressline={Ohio University}, 
            city={Athens},
            postcode={45701}, 
            state={OH},
            country={USA}}

\author[LANL]{R. Tutchton}
\ead{rtutchton@lanl.gov}

\date{\today}

\begin{abstract}
The impact of irradiation on the thermal and electronic properties of materials is a persistent puzzle, particularly defect formation at the atomic and nano- scales. This work examines the nanoscale effects of low-energy irradiation on tungsten (W), focusing on defect-induced modifications to thermal and electronic transport. Using the Site-Projected Thermal Conductivity (SPTC) method [A. Gautam \textit{et al.} PSS-RRL, 2400306, 2024], we analyze bulk and twin-grain boundary W with vacancy defects based on the Norgett-Robinson-Torrens displacements per atom (NRT-dpa) model. SPTC provides a detailed prediction of post-cascade spatial thermal conductivity distribution. We estimate electronic conductivity activity using the "$N^2$ method" [K. Nepal \textit{et al.} Carbon, 119711, 2025] to explore the consequences of vacancies and grain boundaries, highlighting the defect-dependent nature of charge transport behavior. These findings offer high-resolution insights into irradiation-driven transport phenomena, with implications for space-exposed materials and nanoscale thermal/electronic management.
\end{abstract}

\maketitle


\section{\label{sec:introduction}Introduction}

Tungsten (W) is a critical structural material for systems exposed to extreme conditions, such as spacecraft and fusion reactors, due to its remarkable properties. W exhibits high density, the highest melting point of any elemental metal ($\sim$ 3,422$^\circ$C), exceptional resistance to erosion and sputtering, and outstanding radiation tolerance \cite{Wprop1, Wprop2}. In space materials, W is used for components that need to withstand intense heat and mechanical stress, such as rocket engine nozzles, heat and radiation shields (for electronics and crew modules), gyroscope rotors, and inertial guidance system parts within spacecraft \cite{WinSPace, Wprop3,Wprop4}. In low earth orbit (LEO), several types of energetic particles can induce lattice displacement in W by transferring sufficient energy to a single or multiple atom(s). These particles originate from various sources, including the solar wind, cosmic rays, and Earth's magnetosphere as summarized in Table \ref{tab:LEO_particles}. Although  UV radiation ($<$ 300 eV) and atomic oxygen ($\sim$ 5 eV) bombardment does not have exceed the displacement energy of W ($E_d$ $\approx$ 50 to 90 eV \cite{SRIM, PKAreview}), they contribute to other surface erosion process, like oxidation or surface roughening over prolonged exposure. The evolution of these defect structures alters the microstructure, potentially causing embrittlement and impairing the material's thermal and electrical properties. 

\begin{table*}[!t]
    \centering
    \caption{Sources of Energetic Particles in Low Earth Orbit (LEO).}
    \label{tab:LEO_particles}
    \resizebox{\linewidth}{!}{
    \begin{tabular}{|l|l|l|}
        \hline
        \textbf{Source} & \textbf{Particle Types} & \textbf{Energy Range} \\ \hline
        Solar Wind & Protons, electrons, alpha particles, heavy ions & $\sim$1 to 10 keV (typical); $\sim$100 keV during solar storms \cite{t1, t12} \\ \hline
        Solar Energetic Particles & Protons, alpha particles, heavier ions (e.g., Fe) & $\sim$1 GeV during solar flares or coronal mass ejections \cite{t12, t2}\\ \hline
        Galactic Cosmic Rays & Protons, alpha particles, heavy ions (C, O, Fe) & MeV to GeV range, with some exceeding 10 GeV \cite{t3i,t3ii}\\ \hline
        Trapped Radiation (Van Allen Belts) & Protons, electrons & Protons: $\sim$10 to 100 MeV; Electrons: $\sim$1 keV to 10 MeV \cite{t12,t4i,t4ii}\\ \hline
    \end{tabular}
    }
\end{table*}

Modeling the evolution of irradiation-induced damage in extreme space environments presents significant challenges, particularly in understanding its impact on material properties such as thermal and electronic conductivity during irradiation events. Molecular dynamics (MD) simulations have proven invaluable for investigating these phenomena at the atomic scale, enabling detailed insights into system evolution under irradiation \cite{wu2023simulation, PKAsimpaper1, PKAreview, Zarkadoula_2015, W_GAP1, W_GAP2}.  However, most of these studies have focused on defect production statistics and their correlation with the energy of primary knock-on atoms (PKAs), as well as the development of accurate potentials adequate to capture the underlying irradiation mechanisms. While these efforts represent critical advances, there has been minimal exploration of how material properties, such as thermal and electronic conductivity, evolve during or after irradiation events. This gap is understandable, as large system sizes required to accurately model damage and electronic structure computation have been a formidable challenge, until now.

In this work, we predict the atomic thermal conduction activity and electronic active path in W subjected to low energy irradiation. Rather than focusing on the atomistic trajectory during irradiation, we consider the longer term effects of vacancy creation in bulk W as well as vacancies in the vicinity of a twin grain boundary. Irradiation induces diverse phenomena in solids due to the metastable nature of the resulting changes, which depend on the material's structure, interstitial site availability, and other properties. Partial recovery often occurs during or after irradiation, with temperature significantly influencing the structural healing. This study focuses on direct momentum transfer from PKAs, where displaced atoms relax near their original lattice sites, leaving single or double vacancies in the bulk and minor surface disorder.

Our method for predicting thermally active sites in materials, called "Site Projected Thermal Conductivity" (SPTC), was first reported in Reference \cite{SPTC} and follows from the work by Allen and Feldman (AF) on calculations of thermal conductivity \cite{AF_TC}. While the original AF method answers the question: "Given a structure and a force-constant matrix, what is the conductivity tensor?", SPTC extends this by enabling the decomposition of global conductivity into spatially localized estimates of thermal activity. For defect-induced changes in electronic conduction, we utilize the "N-Squared ($N^2$)" method which has been applied to predict and analyze electronic conductivity in copper–carbon composite materials \cite{N2}. The $N^2$ method assumes that electronic conduction paths in materials can be determined by projecting the square of the electronic density of states around the Fermi level onto spatial grids. Both SPTC and the $N^2$ method share the advantage of modest computational cost, making them suitable for simulations involving large supercells. This efficiency is particularly beneficial for studying material properties under radiation-induced defects where large system sizes are unavoidable. 

\section{Computational Details}\label{sec:methods}
Monocrystalline models of body-centered cubic (bcc) tungsten (W) were constructed with orientations  \textbf{x}$||[11\bar{2}]$, \textbf{y}$||[111]$, and \textbf{z}$||[\bar{1}10]$.  These orientations were chosen arbitrarily but designed to align with twin boundary models that feature a mirror plane normal to \textbf{y}$||[111]$. This approach minimizes variables, enabling straightforward comparisons across different models. 

To quantify the number of displaced atoms, $N_d$, for a given damage energy ($T_d$), we adopt the Norgett-Robinson-Torrens displacements per atom (NRT-dpa) model, given as \cite{Kinchin, NORGETT}:

\[
N_d(T_d) = \begin{cases}
0 & T_d < E_d \\
1 & E_d < T_d < 2.5E_d \\
\frac{2.5T_d}{E_d} & 2.5E_d < T_d < \infty
\end{cases}
\]

\noindent where the threshold displacement energy for Tungsten ($E_d$) is taken as 90 eV \cite{SRIM, PKAreview}. Since the LEO energetic particles have $E_d$ $\leq$ 10keV, the NRT-dpa model is suitable and aligns with the more accurate athermal recombination corrected dpa (arc-dpa) proposed by Nordlund \textit{et. al.} \cite{nordlund2018}. Since the arc-dpa confirms that following the instance of a cascade almost all atoms regain positions in the perfect lattice sites (even after thermal spikes \cite{thermalspikes}), we assume that irradiation and subsequent relaxation primarily result in the creation of a single vacancy or (at most) double vacancies in the system, with the rest of the atoms presumed to return to their original lattice positions after cascade relaxation. 

Defect-free models are labeled as "Bulk" and "TGB" for bulk and twin grain boundary structures, respectively. Vacancy configurations are introduced by randomly removing atoms, designated as follows: single vacancy (1SV), two separate single vacancies (2SV), and a double vacancy pair (DV). For the TGB models the vacancies were created close to the grain boundary plane. Surface models were also created by adding a 15 \AA~ vacuum in the z-direction of the Bulk models. Table \ref{tab:models} summarizes the model names and sizes used in the thermal (SPTC) and electronic ($N^2$) activity analyses. 

Initial atomic configurations were generated using ATOMSK \cite{ATOMSK}. The SPTC models were structurally optimized in LAMMPS \cite{lammps} using the conjugate gradient algorithm with the Machine-learning Gaussian Approximation inter-atomic Potential (GAP) for tungsten \cite{W_GAP1, W_GAP2}. $N^2$ models were similarly optimized in VASP \cite{VASP1996} with the Perdew-Burke-Ernzerhof ultrasoft psuedo-potential (US-PBE) \cite{perdew1996}.

\begin{table}[!h]
\centering
\caption{Description of the W models considered in this study. The model names``Bulk" and ``TGB" rrepresent pristine bulk and twin grain boundary models of W, respectively. The suffixes ``1SV" and``2SV"  denote models with a one and two (separated) vacancy sites, while ``DV" indicates a double vacancy pair. The columns labeled ``SPTC" and ``$N^2$" corresponds to the number of atoms in the models used for the SPTC and $N^2$ analyses, respectively.}
\label{tab:models}
\resizebox{\columnwidth}{!}{%
\begin{tabular}{|c|c|c||c|c|c|c|}
 \hline
    \rowcolor{headergray}  & \multicolumn{2}{c||}{\color{white}\textbf{System Size}} &  & \multicolumn{2}{c|}{\color{white}\textbf{System Size}} \\ \hline
    \rowcolor{headergray} \color{white}\textbf{Model Name} &  SPTC & \textbf{{$N^2$}} & \color{white}\textbf{Model Name} & SPTC & \textbf{{$N^2$}}  \\ \hline
    \rowcolors{3}{lightgray}{darkgray}
    Bulk & 2000 & 270 & TGB & 1800 & 288 \\ \hline
    Bulk1SV & 1999 & 269 & TGB1SV & 1799 & 287  \\ \hline
    Bulk2SV & 1998 & 268 & TGB2SV & 1798 & 286 \\ \hline
    BulkDV & 1998 & 268 & TGBDV & 1798 & 286 \\ \hline
\end{tabular}%
}
\end{table}

\begin{figure*}[!t]
 \centering
    \includegraphics[width=\linewidth]{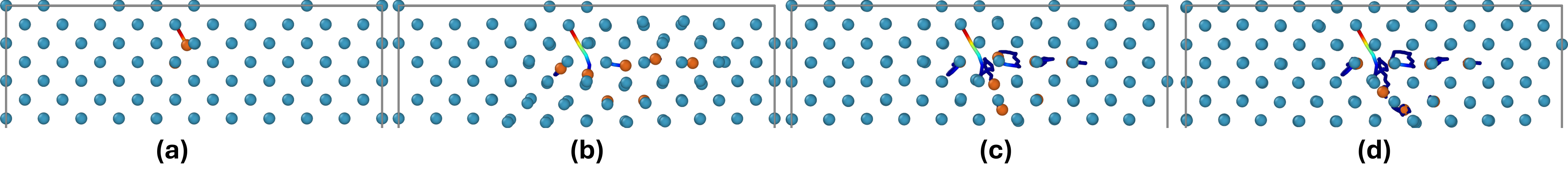}
    \caption{Trajectory of a single PKA: (a) It gains energy and is ejected from its lattice site. (b) Energy transfer agitates neighboring atoms. (c) Most atoms return to their sites, but (d) the PKA loses its original position and becomes an interstitial, leaving a vacancy (Frenkel defect).}
    \label{fig:cfig_traj}
\end{figure*}

\begin{figure*}[!t]
 \centering
    \includegraphics[width=\linewidth]{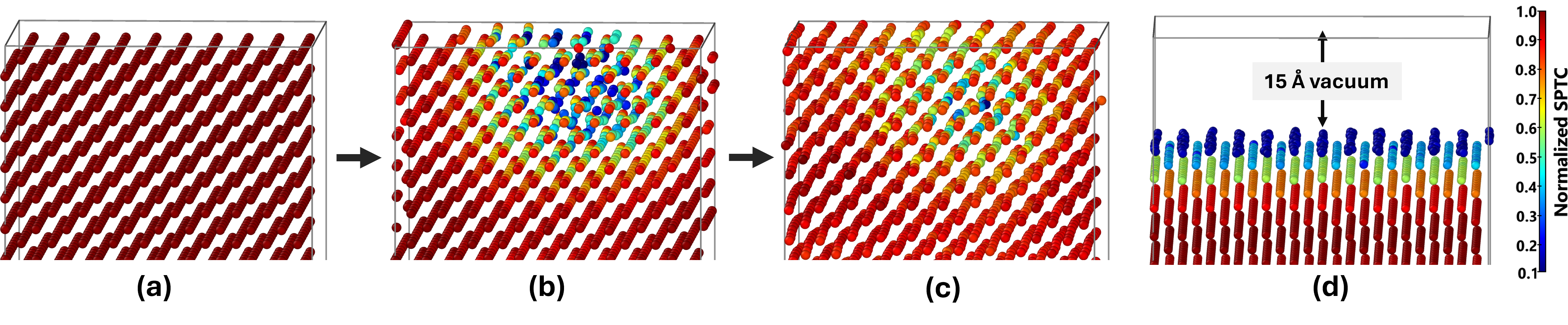}
    \caption{SPTC monitors thermal transport changes during a multi-PKA cascade. (a) Shows the unperturbed lattice, (b) depicts site displacements, and (c) presents the system just before full cascade relaxation. (d) The surface model highlights SPTC's ability to detect surface disturbances from low-energy particles (< 200 eV) that don't create vacancies. The colorbar indicates SPTC values normalized to the unperturbed lattice maximum in (a).}
    \label{fig:cfig_SPTCtraj}
\end{figure*}

Details of the SPTC and $N^2$ methods are available in References \cite{SPTC} and \cite{N2}, respectively. However, we provide additional discussion on the underlying logic of these methods in \ref{app:SPTC} (SPTC) and \ref{app:N2} ($N^2$). The $N^2$ method was implemented using static calculations from VASP \cite{VASP1996} with the US-PBE pseudo-potential at a single \textbf{k}-point ($\Gamma$). A fixed plane-wave energy cutoff of 520 eV was applied across all calculations. The Kohn-Sham orbitals were computed on a $60 \times 60 \times 60$ grid, and 30 band states above and below the Fermi level (60 in total) were included in the calculations.

Computing the SPTC requires first obtaining the dynamical matrix, $D^{\alpha \beta}_{i j}$, which relates to the force constant, $\phi_{x,x^\prime}^{\alpha \beta}(0, \gamma)$,  as:

\begin{subequations}\label{eq:DM_FC}
\begin{gather}
    D^{\alpha \beta}_{x x^\prime} = \frac{1}{\sqrt{m_x m_{x^\prime}}} \phi_{x,x^\prime}^{\alpha \beta}(0, \gamma) \\
    \phi_{x x^\prime}^{\alpha \beta}(\gamma^\prime,\gamma) =  \frac{\partial^2E}{\partial \alpha_{\gamma^\prime x} \partial \beta_{\gamma x^\prime}}.
\end{gather}
\end{subequations}

\noindent where E is the potential energy,  $\alpha_{\gamma x}$ is the $\alpha$-component of the displacement vector for the atom (of mass $m_x$) in site $x$ of the $\gamma^{\text{th}}$ cell. The dynamical matrix was obtained using GAP potential in LAMMPS by displacing all atoms in the supercell by 0.015 \AA{} in all Cartesian directions. The vibrational frequency, $\omega_m$, of the $m^\text{th}$ normal mode can be determined by solving the eigenvalue problem for the classical normal modes, defined for the center of the phonon Brillouin zone for \textbf{k=0} as: 

\begin{equation}\label{eq:freq}
        \omega^2_m \  e^{\alpha, m}_{x, \textbf{0}} = \sum_{\beta j} D^{\alpha \beta}_{x x^\prime} e^{\beta, m}_{x^\prime, \textbf{0}}
\end{equation}

\noindent Here, $e^{\alpha, m}_{x, \textbf{k}}$ is the polarization of the $m^\text{th}$ normal mode for the atom in site $x$ along the $\alpha$ direction.  The vibrational density of states (VDoS), $g_h(\omega)$, is then easily obtained from a Gaussian density estimate of the frequencies as: 

\begin{equation}\label{eq:VDoS}
    g_h(\omega) = \frac{1}{Nh\sigma}\frac{1}{\sqrt{2\pi}}\sum_{i = 1}^{3N}\exp{\left[\frac{-(\omega - \omega_i)^2}{2h^2\sigma^2}\right]}
\end{equation}

\noindent N is the number of atoms and $\omega_i$ is the contribution of the $i^\text{th}$ atom to the eigenfrequency in a given Cartesian direction.  $\sigma$ is the standard deviation of the distribution and $h$ is the kernel (smoothing) bandwidth calculated as \cite{silverman}:

\begin{equation}\label{eq:h}
    h = 0.9~ \text{min}\left(\sigma, \frac{\text{IQR}}{1.34}\right)N^{-0.2}
\end{equation}

\noindent where IQR is the interquartile range of the distribution \cite{IQR}.

\section{\label{sec:RandD} Results and Discussion}

To validate the NRT-dpa prediction, we simulated cascades with both single and multiple PKAs for damage energies ranging from 100 eV to 500 eV on the Bulk W model. One of such trajectories, depicted in Figure \ref{fig:cfig_traj} (a–d), shows the creation of a vacancy as a 300 eV PKA moves along its path. During its motion through the bulk, the PKA displaces neighboring atoms, temporarily disturbing the lattice. However, as shown in Figure \ref{fig:cfig_traj} (b–d), most of these displaced atoms eventually return to their original lattice sites. These simulations validates our focus on a maximum of double vacancy creation due for the energy range of LEO particles in Table \ref{tab:LEO_particles}. 

To evaluate SPTC's effectiveness in tracking thermal activity during a cascade, we initiated a multi-PKA cascade in both the bulk and surface models. Figures \ref{fig:cfig_SPTCtraj}a -- c illustrate its ability to capture thermal variations throughout the cascade. The colorbar represents SPTC values normalized to the unperturbed lattice maximum in Figure \ref{fig:cfig_SPTCtraj}a, where all atoms initially exhibit the same maximum SPTC value. However, as the cascade progresses, Figure \ref{fig:cfig_SPTCtraj}b reveals changes in SPTC, which persist into Figure \ref{fig:cfig_SPTCtraj}c as the system relaxes post-cascade. Notably, SPTC effectively identifies potential vacancy sites by highlighting atoms with the lowest values at the conclusion of one cascade event and the onset of another.

Since the NRT-dpa model suggests that LEO particles with energies below 2\(E_d\) (\(\lessapprox\) 200 eV) are unlikely to cause atomic displacement, such interactions primarily affect the material's surface. This behavior is well captured in the SPTC prediction for the surface model of tungsten in Figure \ref{fig:cfig_SPTCtraj}d, where surface atoms exhibit the lowest (but nonzero) thermal activity after cascade relaxation. This observation aligns with the degradation of W’s thermal response following low-energy radiation \cite{Wdamage1,Wdamage2}. 

\begin{figure}[!t]
 \centering
    \includegraphics[width=\linewidth]{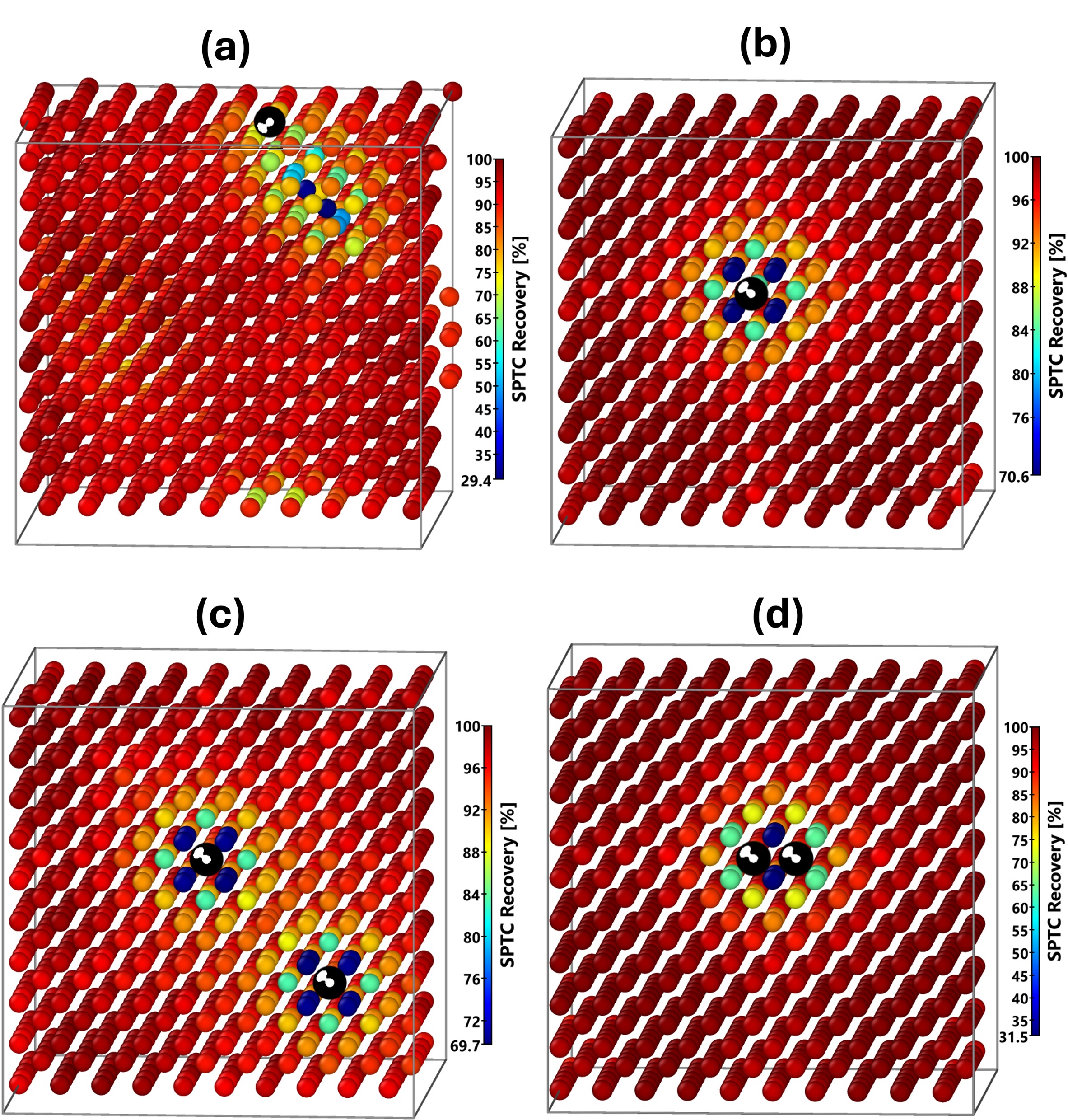}
    \caption{Thermally active sites in bulk W models. The colorbar represents SPTC normalized to its maximum value in the bulk [in \%]. Vacancy sites are shown as black/white spheres. (a) Frenkel defect post-relaxation from Figure \ref{fig:cfig_traj}, while (b), (c), and (d) correspond to the Bulk1SV, Bulk2SV, and BulkDV models, respectively. In (a), the blue sphere closest to the vacancy site represents the interstitial atom while the adjacent blue sphere corresponds to a displaced atom resulting from the presence of the interstitial.}
    \label{fig:cfig_bulkSPTC}
\end{figure}

Next, we predict the thermal behavior of bulk W models containing vacancies. Figure \ref{fig:cfig_bulkSPTC} illustrates the SPTC values for atomic sites in bulk W across different defect configurations: (a) Frenkel defect (post-cascade relaxation), (b) Bulk1SV, (c) Bulk2SV, and (d) BulkDV. Vacancy sites are represented as black/white spheres. The colorbar denotes the normalized thermal conductivity magnitude, ranging from blue (lowest) to red (highest). In Figure \ref{fig:cfig_bulkSPTC}a, the interstitial atom appears as the blue sphere closest to the vacancy site at the top of the supercell, while the adjacent blue sphere corresponds to a displaced atom, due to the presence of the interstitial atom. These atoms exhibit the lowest thermal activity, with a reduction of approximately 71\%. The single vacancy (Bulk1SV) in Figure \ref{fig:cfig_bulkSPTC}b exhibits a thermal activity reduction of  $\approx 30\%$. Similarly, Bulk2SV in Figure \ref{fig:cfig_bulkSPTC}c exhibits a reduction of $\approx 31\%$. Notably, as the spacing between defects decreases, the likelihood of thermal activity recovering to bulk levels diminishes. The thermal response of the double vacancy (BulkDV, Figure \ref{fig:cfig_bulkSPTC}d) closely resembles that from the model with a Frenkel defect in Figure \ref{fig:cfig_bulkSPTC}a, with a significant reduction in thermal activity of about $69.5\%$.

\begin{figure}[!t]
    \centering
    \includegraphics[width=\linewidth]{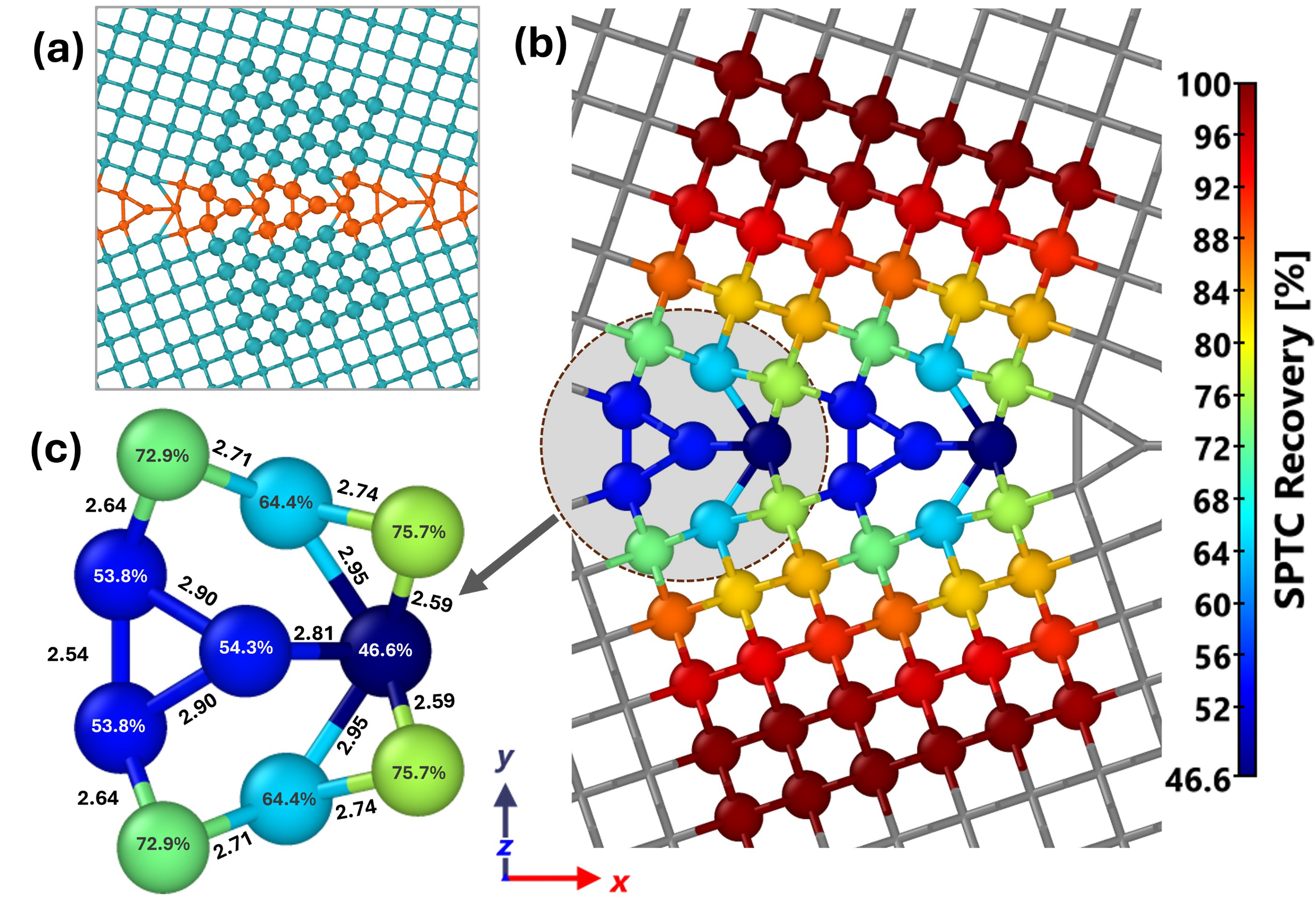}
    \caption{Thermal transport activity in the TGB model. (a) The selected region of interest, where atoms of significance are represented by spheres with a larger radius. The mirror plane atoms are colored in brown, while the remaining atoms are shown in blue (see also Figure S1a). (b) The spatial distribution of SPTC, with a colorbar indicating the percentage recovery of SPTC to its maximum value away from the mirror plane. (c) The recovery values for a repeating mirror plane region (dotted circle with gray overlay), with interatomic distances (in \AA) highlighted to illustrate its correlation to SPTC values.}
    \label{fig:cfig_TGBSPTC}
\end{figure}

Understanding the influence of grain boundaries on thermal activity is crucial for controlling nanoscale thermal transport in materials. The thermal activity in the twin-grain boundary model without vacancies (TGB) is shown in Figure \ref{fig:cfig_TGBSPTC}, with a focus on the mirror plane region of the lattice. The selected lattice region is displayed in Figure \ref{fig:cfig_TGBSPTC}a, where the atoms of interest are highlighted with larger spheres. The mirror plane atoms are colored in brown, while the rest are shown in blue. The thermal activity distribution obtained from SPTC calculations for this region is presented in Figure \ref{fig:cfig_TGBSPTC}b, with the colorbar representation consistent with the Bulk models in Figure \ref{fig:cfig_bulkSPTC}. The SPTC distribution for the entire supercell is also provided in Figure S1(a) of the supplementary material.

Figures \ref{fig:cfig_TGBSPTC}b and S1(a) indicate that the lowest SPTC values occur at the mirror plane atoms, increasing outward in an almost uniform pattern on either side of the plane. This trend is expected, as the lattice retains its crystalline periodicity away from the grain boundary. Unlike bulk models with a uniform lattice parameter, the introduction of grain boundaries distorts atomic positions and alters interatomic spacing in the boundary region. Figure \ref{fig:cfig_TGBSPTC}c provides a magnified view of the highlighted region (dotted circle with gray overlay), revealing that these distortions in atomic spacing (in \AA) at the grain boundary significantly influence thermal activity. The atom with the deepest blue color, corresponding to the lowest SPTC value (46.6\%), also exhibits the greatest variation in interatomic distances, followed by sites with SPTC values of 54.3\%, 53.8\%, 64.4\%, and so on. A similar thermal activity pattern has been observed for silicon twin grain boundaries \cite{SPTC}. Therefore, at the macroscale, grain boundaries are known to reduce thermal conductivity by disrupting collective atomic motion within the periodic lattice, thereby enhancing phonon scattering \cite{phonon1, phonon2, phonon3}. However, at the nanoscale, local variations in interatomic spacing near the grain boundary region can also introduce significant changes in thermal activity, highlighting an additional mechanism influencing thermal transport in materials with grain boundaries.

\begin{figure}[!t]
    \centering
    \includegraphics[width=\linewidth]{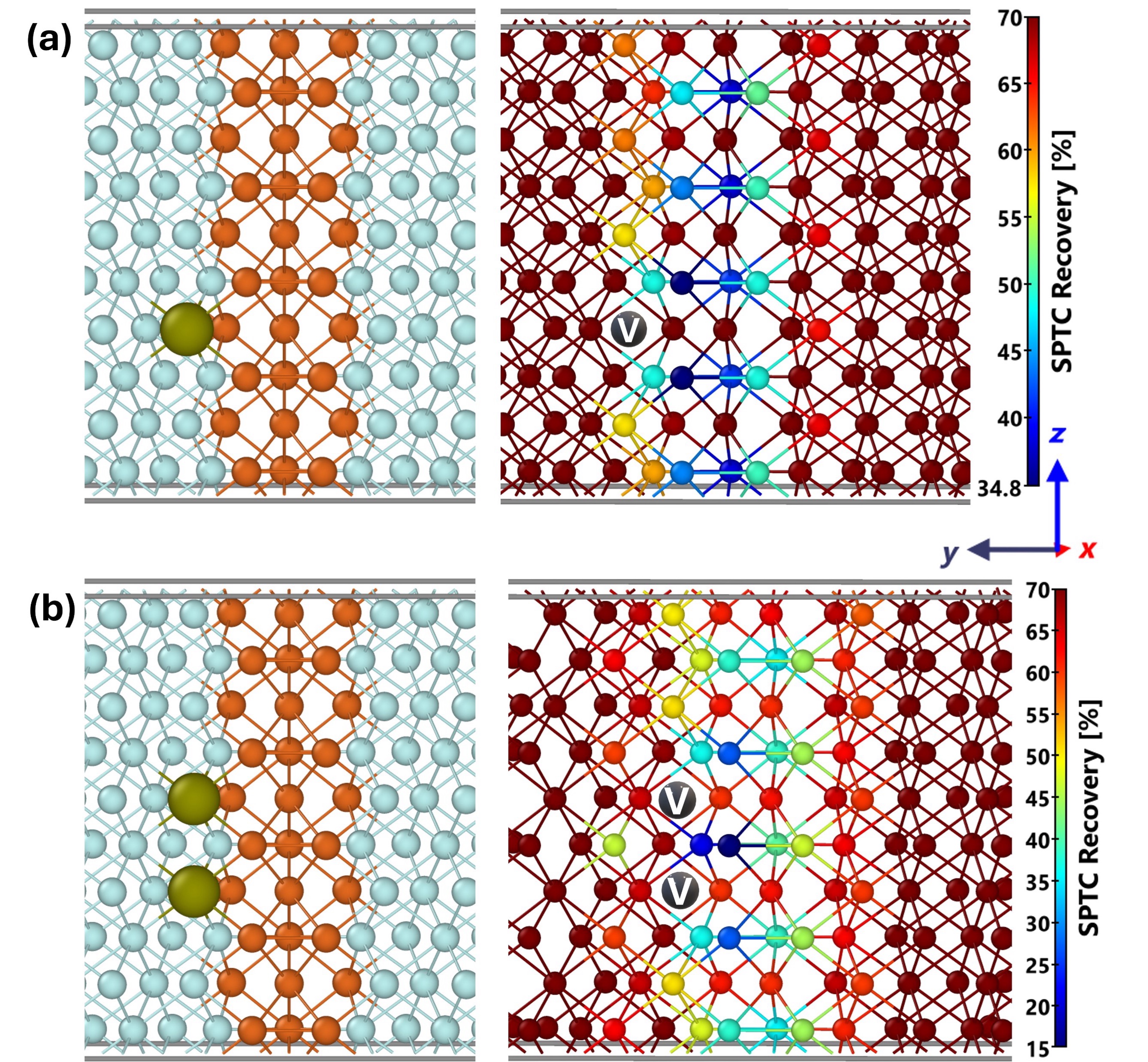}
    \caption{Thermally active sites in (a) TGB1SV and (b) TGBDV models. The left column provides a reference for identifying the vacancy site(olive-drab spheres), while the right column presents the SPTC distribution. The colorbar in the right column, representing SPTC recovery, is clipped at 70\% to emphasize the specified region. The black spheres labeled with "V" denotes the vacancy sites}
    \label{fig:cfig_TGBVacSPTC}
\end{figure}

\begin{figure}[!b]
    \centering
    \includegraphics[width=\linewidth]{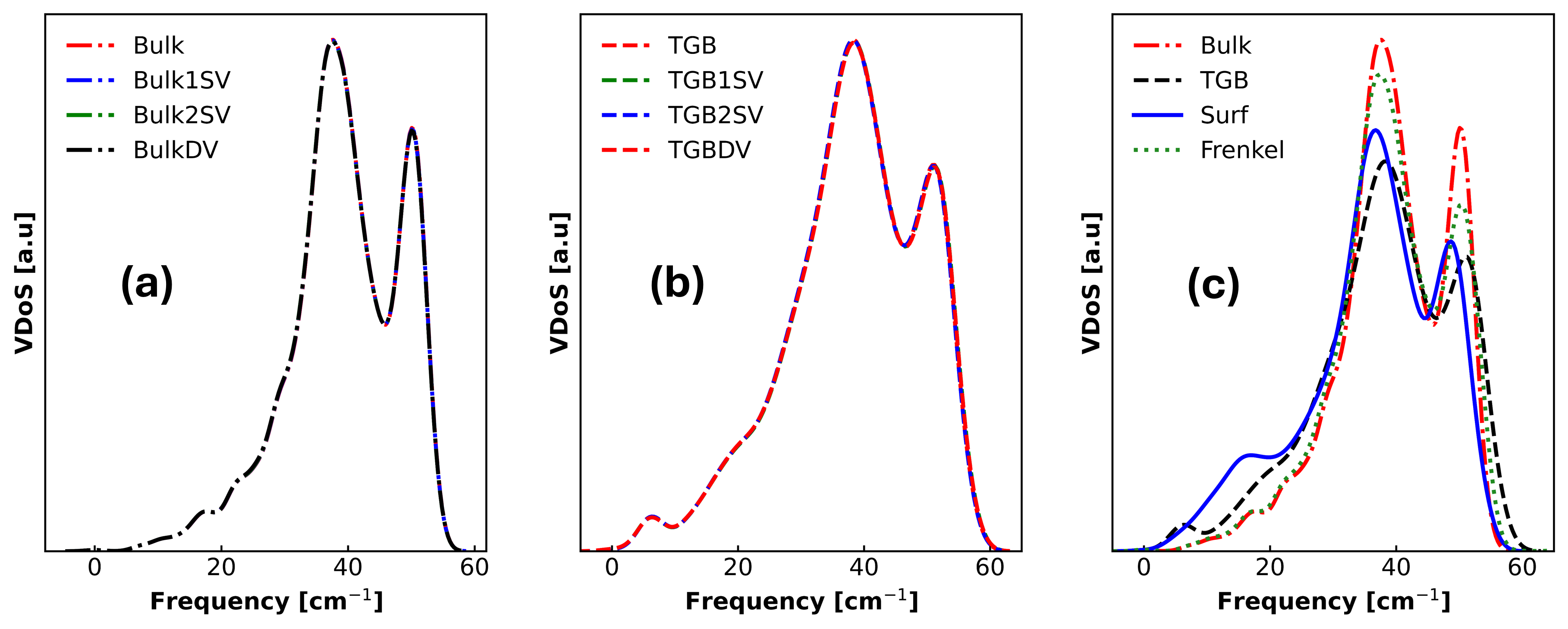}
    \caption{Comparison of vibrational density of states for W models. No difference is observed from vacancy creation in (a) bulk or (b) twin grain boundary models. (c) While the overall intensity vary across W systems, peak positions remain largely unchanged.}
    \label{fig:cfig_VDoS}
\end{figure}
 
The inclusion of vacancies in the vicinity of the grain boundary region further influences thermal conduction activity. Figure \ref{fig:cfig_TGBVacSPTC}a and b present the SPTC distribution around the grain boundary region for TGB1SV and TGBDV, respectively. The left column is a guide for identifying the vacancy sites (olive-drab spheres) and the mirror plane atoms (brown spheres), while the right column illustrates the corresponding SPTC distribution. In the right column, the black spheres labeled with "V" also denote the vacancy sites. To emphasize on this region the SPTC has been cutoff at a maximum of 70\%, however the full SPTC distribution for TGB1SV and TGB2SV are shown in Figure S1(b) and (c), respectively. Hear the deepest red color of atoms in the right column of Figure \ref{fig:cfig_TGBVacSPTC} have SPTC above 70\%.

The SPTC value due to vacancy sites decreased to 34.8\% in TGB1SV, however, this is still a considerable  75\% of the minimum value in the mirror plane in the TGB model. In contrast, the SPTC in TGBDV is significantly reduced to just 15\% of the bulk maximum. A visual comparison reveals that the introduction of a single vacancy in the transition from TGB $\rightarrow$ TGB1SV $\rightarrow$ TGBDV leads to a substantial alteration in the SPTC distribution pattern. Unlike in TGB, where the SPTC recovery pattern remains nearly uniform in both directions from the mirror plane, this uniformity is disrupted in TGB1SV and TGBDV. Specifically, in TGB1SV, a greater number of atoms in the mirror plane exhibit SPTC values exceeding 70\% compared to TGBDV. Moreover, the additional vacancy in TGBDV further impedes SPTC recovery, extending up to two lattice sites beyond the vacancy location. 

It is worth noting that while thermal property estimates are typically derived from global averaging of the system, the local probing capability of SPTC provides valuable insights. For instance, the VDoS shown in Figure \ref{fig:cfig_VDoS}, obtained from the normal vibrational modes using Equation \ref{eq:VDoS}, exhibits no discernible differences among the various configurations within the Bulk or TGB models (Figures \ref{fig:cfig_VDoS}a and \ref{fig:cfig_VDoS}b, respectively). Furthermore, while subtle variations are present in the vibrational spectra across different systems—including Bulk, TGB, surface models, and models with Frenkel defects—the peak positions remain largely unchanged.

\begin{figure}[!h]
    \centering
    \includegraphics[width=\linewidth]{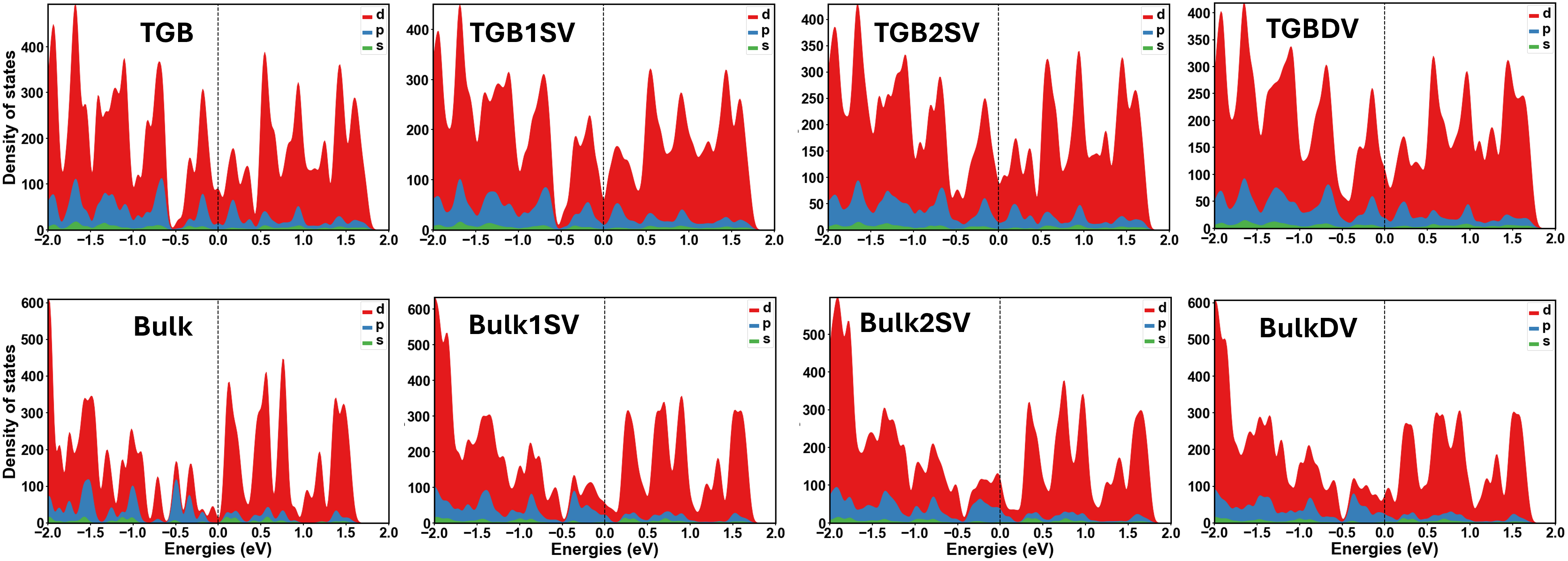}
    \caption{Orbital-projected electronic density of states for W models: \(s\) (green), \(p\) (blue), and \(d\) (red) orbitals. the dashed black line indicated the Fermi energy (shifted to 0).}
    \label{fig:cfig_EDoS}
\end{figure}

Similar to the vibrational density of states, the orbital-projected electronic density of states (EDoS) also fails to capture the impact of a small number of vacancies on electronic conductivity. This limitation is illustrated in Figure \ref{fig:cfig_EDoS}, which presents the contributions of the \(s\) (green), \(p\) (blue), and \(d\) (red) orbitals for both the twin grain boundary (top) and bulk (bottom) models. In contrast, the \(N^2\) method, akin to the space-projected (electronic) conductivity (SPC) approach \cite{SPC1, SPC2, SiOx, kishor1}, provides a spatially resolved description of electronic conduction, making it more effective in identifying localized defect-induced changes in conductivity.

\begin{figure*}[!t]
 \centering
    \includegraphics[width=\linewidth]{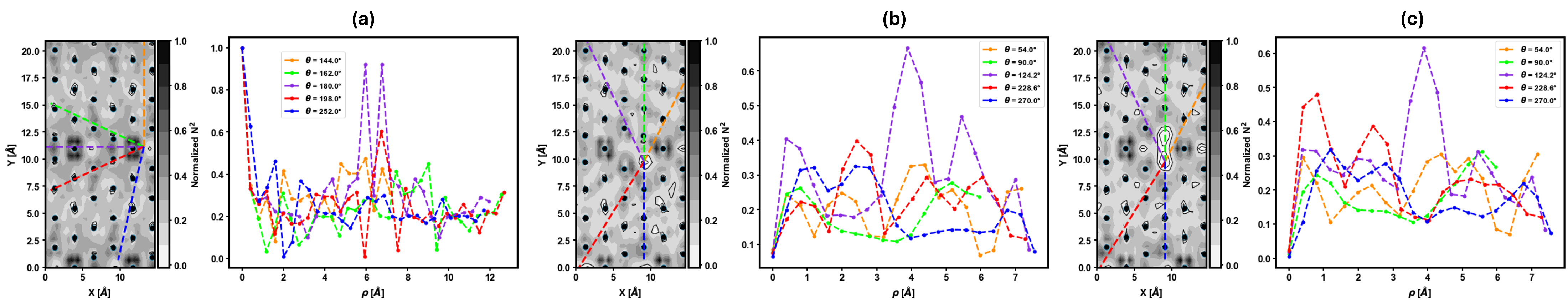}
    \caption{Variation in electronic conductivity along the radial directions from a point in (a) TGB, (b) TGB1SV, and (c) TGB2SV. The 2D grayscale plots on the left represent cross-sections at \( Z \approx 6.71 \) \AA. Atoms are depicted as black spheres, while conductivity magnitude increases from white to black. The colored dashed lines indicate the radial paths (\(\rho\)) from a point at different angles (\(\theta\)), along which \( N^2 \) values are extracted and plotted on the right.}
    \label{fig:cfig_TGBN2Radial}
\end{figure*}

\begin{figure*}[!t]
 \centering
    \includegraphics[width=\linewidth]{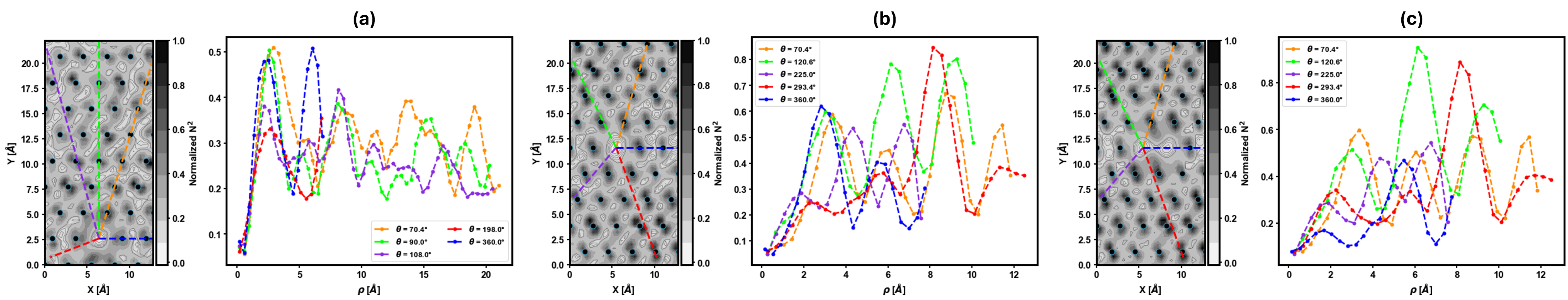}
    \caption{Variation in electronic conductivity along the radial directions from (a, b) single and (c) double vacancy sites. The 2D grayscale plots on the left represent cross-sections at (a) \( Z \approx 4.48 \) \AA ~ and (b, c)  \( Z = 6.71 \) \AA. Atoms are depicted as black spheres, while conductivity magnitude increases from white to black. The colored dashed lines indicate the radial paths (\(\rho\)) from a vacancy site at different angles (\(\theta\)), along which \( N^2 \) values are extracted and plotted on the right}
    \label{fig:cfig_BULKN2Radial}
\end{figure*}

The $N^2$ electronic conduction paths for the W twin grain boundary models are shown in Figure S2 for TGB, TGB1SV, TGB2SV, and TGBDV in rows I, II, III, and IV, respectively. Although electronic conduction extends across the entire system (unlike SPTC which is localized on atomic sites), we present projections at atomic planes located at $Z \approx$ 2.24, 4.48, 6.71, 8.95, and 11.19~\AA~ in five columns. Atomic sites are represented by blue circles, while white arrows mark vacancy sites. 

Generally for the TGB model, the electronic conduction remains consistent across all five planes (a--e), with the highest activity localized in the mirror plane, forming a "dome-like" glow along the y-axis. A small, diamond-shaped conduction path appears between the pair of atoms closest to the mirror plane, indicated by the white arrow in Figure S2 [I](c). The electronic conductivity is anisotropic and this is illustrated from the radial projection of the conductivity in the Z $\approx$ 6.71~\AA~ atomic plane for TGB in Figure \ref{fig:cfig_TGBN2Radial}a. The colored lines in the grayscale plot show the radial path, $\rho$ (at different angles, $\theta$ from a point), along which the conductivity is measured and plotted on the right. Similar plots are shown in Figure \ref{fig:cfig_TGBN2Radial}b and c for TGB1SV and TGB2SV, respectively. The variation in the conductivity plot is more pronounced along the grain boundary with maximum values in the vicinity of the mirror-plane atoms. 

For TGB1SV in Figure \ref{fig:cfig_TGBN2Radial}b and Figure S2 [II](c), the creation of a vacancy site modifies the conduction distribution, reorienting the dome-like glow towards the x-axis and forming a new conduction path between the mirror plane atom and its neighboring site. Further modifications occur in the TGB2SV (Figure \ref{fig:cfig_TGBN2Radial}c and S2 [III](c)), where the conduction pattern becomes more localized near the mirror plane atom closest to the vacancies. In Figure \ref{fig:cfig_TGBN2Radial}b and c, the radially projected electronic conductivity distributions are unique for each angle $\theta$. For example,  the second most-promient peak for the purple line, which is at $\theta = 124.2^\circ$ and $\rho \approx 5.8$ \AA, is significantly smaller in TGB2SV due to its extra vacancy site. Clearly, the inclusion of the second vacancy site disrupts the new conduction path observed in the TGB1SV model, further altering the electronic transport characteristics. 

The TGBDV model in Figure S2 [IV] shows a conductivity pattern that suggests a strong dependence of conductivity distribution on the vacancy location. Here the second vacancy is positioned in the same atom cluster below the mirror plane. It is observed that the strong glow in the Bulk model also changes orientation (as discussed for TGB1SV)  in Figure S2 [IV](b). Also, the mirror plane atom forms a new conduction path with the atom closest to the vacancy, as observed in Figure S2 [II](c). This strongly indicates that vacancy creation affects the conductivity distribution for atoms in the grain boundaries.

In general, the distribution pattern of the undisturbed lattice sites appears slightly altered due to the presence of the vacancies. This becomes increasingly obvious in the $N^2$ plot for the bulk models in Figure S3, depicted for the Bulk, Bulk1SV, Bulk2SV, and BulkDV models in rows I, II, III, and IV, respectively. Again only planes with atom sites are shown for Z $\approx$ 2.24, 4.48, 6.71, 8.95, and 11.19 \AA~ are shown. The distribution pattern for the electronic conduction in the Bulk model Figure S3 [I] (a -- e) appears in repeating patterns since the atoms are in their equilibrium lattice sites. We note that the pattern is due to the crystallographic orientation chosen for the bulk models which is: \textbf{x}$||[11\bar{2}]$, \textbf{y}$||[111]$, and \textbf{z}$||[\bar{1}10]$. 

Bulk1SV and Bulk2SV in Figure S3 [II](c) and [III](c) both exhibit significant modifications in conductivity patterns across all planes, particularly where the vacancy is located. However, Bulk2SV displays a more irregular and dispersed conductivity pattern compared to Bulk1SV. Moreover, unlike in Bulk1SV, the influence of the vacancy sites in Figure S3 [III](b) and (d) extends to the preceding planes (a) and (c), respectively, indicating a broader spatial impact on electronic conduction. The electronic conduction active path in the bulk models with vacancies form dumbbell-shaped distributions around the atomic sites. Similar to the twin grain boundary models, the introduction of a single vacancy in a bulk system induces an anisotropic electronic conductivity distribution. Figures \ref{fig:cfig_BULKN2Radial}a and \ref{fig:cfig_BULKN2Radial}b illustrate the radially projected conductivity for two atomic planes containing single vacancy sites at $Z \approx 4.48$ \AA \ and $Z \approx 6.71$ \AA, respectively. In Figure \ref{fig:cfig_BULKN2Radial}a, the first conductivity peaks for all angles are consistently positioned around $\rho \approx 3$ \AA. However, in Figure \ref{fig:cfig_BULKN2Radial}b, the first peak positions exhibit greater variation, indicating that defect location plays a crucial role in shaping the electronic conductivity distribution in W.

The electronic conductivity distribution near the divacancy in BulkDV, shown in Figure S3 [IV](c), also exhibits a localized dumbbell-shaped pattern around the atoms and is similar to the distribution observed in aluminum by Subedi \textit{et al.} using the SPC method \cite{AlSPC}. The conductivity distribution in other planes closely resembles that of the Bulk1SV model, as the second vacancy was intentionally introduced adjacent to the vacancy site in Bulk1SV. However, the conductivity distribution around the vacancy appears more diffused, suggesting that the presence of an additional vacancy further alters the electronic conduction pathways in bulk W. This effect is further evident in the radial projection of the conductivity pattern associated with the double vacancy sites, as shown in Figure \ref{fig:cfig_BULKN2Radial}c for the $Z \approx 6.71$ \AA~ plane. The previously noted similarity between the Bulk1SV and BulkDV models is further confirmed in the radial projection at all angles, except at $\theta = 360^\circ$, where the conduction path directly intersects both vacancy sites.

\section{\label{sec:conclucion} Conclusion}

This study investigates the atomic-level behavior of tungsten (W) subjected to low- to mid-energy irradiation, akin to that induced by energetic particles in low Earth orbit. The primary objective is to understand how irradiation-induced defects, including vacancies and interstitials, influence the distribution of thermal and electronic conductivity in tungsten at the nanoscale.

The analysis considered both bulk and twin grain boundary structures of W, with defects introduced according to the Norgett-Robinson-Torrens (NRT) displacements per atom (dpa) model. This was achieved by artificially incorporating single and double vacancies at specific locations within the W models. Thermal conductivity was examined using the Site-Projected Thermal Conductivity (SPTC) method, a novel technique that effectively tracks lattice disturbances caused by energetic primary knock-on atoms (PKAs) and predicts the thermal conductivity distribution in W following cascade relaxation. A notable correlation was observed between the relaxed atomic distances and the thermal conductivity variations near the grain boundaries in the twin-grain boundary models.

The electronic conductivity distribution was analyzed using the $N^2$ method, revealing that vacancies and grain boundaries substantially modify the electronic conduction pathways. The spatial conductivity pattern was found to depend on the type and location of vacancies—whether single or double, and situated in the bulk or near a grain boundary. Furthermore, the presence of vacancies introduces anisotropy into the electronic conductivity distribution, demonstrating a path-dependent nature.

In conclusion, this work provides insights into the nuanced changes in thermal and electronic conductivity distributions in large supercells of tungsten, which are not readily apparent through conventional approaches such as vibrational and electronic density of states analysis. It further highlights that even low vacancy concentrations in tungsten can induce significant changes in its material properties at the nanoscale, hence, highlighting the importance of atomic-level defect analysis for understanding radiation-induced property modifications. 

\section*{Acknowledgment}  
C. U. thanks Aashish Gautam and Kishor Nepal for insightful discussions that were useful to this work. Simulations were performed with resources provided by the Los Alamos National Laboratory Institutional Computing Program and the National Energy Research Scientific Computing Center (NERSC), a Department of Energy Office of Science User Facility.

\section*{Funding}
This work was supported by the U.S. Department of Energy through the Los Alamos National Laboratory. Los Alamos National Laboratory is operated by Triad National Security, LLC, for the National Nuclear Security Administration of U.S. Department of Energy (Contract No. 89233218CNA000001). Research presented in this article was supported by the Laboratory Directed Research and Development program of Los Alamos National Laboratory under the Director's Postdoctoral Fellowship Program, project number 20240877PRD4. 

\appendix

\section{Site-Projected Thermal Conductivity}\label{app:SPTC}
In the late fifties, Kubo discovered formulae for linear transport coefficients based on what we now call the Fluctuation-Dissipation Theorem, which links fluctuations at equilibrium to processes dissipating energy \cite{Kubo_EC,Kubo_TC}. These expressions involve the temporal autocorrelation functions of currents, for example the electrical or heat current, computed for a system in thermal equilibrium. These “Kubo formulas” have been the main avenue to compute transport from atomistic systems in equilibrium ever since. An intrinsic limitation of all such formulae is that the external perturbation (for example electric field or thermal gradient) is weak enough to justify the assumption of linear response.

In this paper, we consider the special case of thermal transport. For classical dynamics obtained from a molecular dynamics simulation, it is simple to compute the autocorrelation function (ACF) directly from the computed instantaneous particle dynamics with a suitable assumption for the form of the heat current, usually the form due to Hardy \cite{SPTC}. The thermal conductivity is obtained from a quadrature over the full support of the ACF. This is now a standard calculation available in codes like LAMMPS \cite{NPC}. The only care needed for this estimate of the conductivity is that the simulation is run long enough that the ACF vanishes. An attractive feature of this scheme is that anharmonic effects stemming from the selected potential are fully included. The physical validity of the result depends on the accuracy of the potential, the choice of the form for the heat current, and the temperature. For temperatures well below the Debye temperature such calculations are suspect.

Allen and Feldman fixed the last-named limitation imposed by classical lattice dynamics by invoking the harmonic approximation (HA) and working out the conductivity tensor for a quantized lattice, thus producing a theory value for all temperatures for which the HA makes sense \cite{AF_TC}. The computation of the ACF in this theory is gracefully handled with the HA, and results in a formula for the thermal conductivity requiring (1) the force-constant matrix, (2) the eigenvalues and vectors of the dynamical matrix, (3) the coordinates and masses of the atoms and of course information about the periodicity. There have been many calculations of thermal transport using this method which have led to basic insights into transport in homogeneously disordered systems, and the helpful analysis of the energy-dependence of transport — the nature of quantum thermal transport as a function of phonon energy \cite{AFapply}. The method is readily implemented even for \textit{ab initio} interactions (e.g. density functional methods), and so can be applied to diverse materials.

This outstanding body of work offers no insight into one question of special importance for Materials Design: “How do individual atoms contribute to thermal transport?”. This is a key question for disordered systems — what manifestations of local disorder impede or enhance thermal transport? For systems involving interfaces, point and/or extended defects, or homogeneously disordered systems such information is valuable and might suggest improved materials for applications, the ultimate hope being a theoretical prescription to tailor the thermal transport properties.

In this paper we apply a method that offers an answer to the question posed in the preceding paragraph. We have published a series of papers detailed what we have called the “space projected (electrical) conductivity” \cite{SPC1} and obtained highly informative results about local electrical conduction activity in materials. Recently, we extended this concept to thermal transport using a simple strategy. If one explicitly writes out the equations of AF for the conductivity tensor, this involves several nested sums over the normal mode (phonon) frequencies,  various terms involving atomic positions in the cell and lattice translations, the force constant matrix, and the eigenvectors of the dynamical matrix. Since all these sums are finite it is obviously possible to carry the sums out in any order. Thus, we can exactly (in the context of AF) write the TC as \cite{SPTC}:

\begin{equation}\label{sptc}
    \kappa = \sum_{x, x'} \Xi(x, x'),
\end{equation}

\noindent where $\Xi(x,x^\prime)$ at $\textbf{k} = 0$ as \cite{SPTC}:

\begin{align}
\label{eq:Gamma_def} 
   \Xi(x, x') &= \frac{\pi \hbar^2}{48k_BT^2V} \frac{1}{\sqrt{m_{x} m_{x^\prime}}} \sum_{\gamma}   \left(\textbf{R}_{\gamma}  + \textbf{R}_{xx'}\right) \sum_{m, n \neq m} \delta(\omega^{\textbf{0}}_m - \omega^{\textbf{0}}_n -\omega) \notag \\
    &  \frac{(\omega^{\textbf{0}}_m + \omega^{\textbf{0}}_n)^2}{\omega^{\textbf{0}}_m \omega^{\textbf{0}}_n}  \left( \frac{e^{\beta \hbar \omega_m}}{\left(e^{\beta \hbar \omega_m} - 1\right)^2}\right)
    \sum_{\alpha \beta} \phi^{\alpha \beta}_{x, x'}(0, \gamma) e^{\alpha,m}_{x, \textbf{0}} e^{\beta,n}_{x', \textbf{0}}  \notag \\
    &  \sum_{\gamma'} \sum_{y,  y^\prime} \sum_{\alpha^\prime, \beta^\prime}\frac{1}{\sqrt{m_y m_{y^\prime}}}  \phi^{\alpha' \beta'}_{y, y^\prime} (0, \gamma')  e^{\alpha',m \dagger}_{y, \textbf{0}} e^{\beta',n \dagger}_{y^\prime, \textbf{0}} \left(\textbf{R}_{\gamma}  + \textbf{R}_{yy^\prime}\right).
\end{align} 

\noindent $V$ and $T$  are the volume and the temperature (300 K) of the system, $\textbf{R}_\gamma$ is the position vector for the $\gamma^{\text{th}}$ cell, $\textbf{R}^{\eta}_{x,x^\prime}$ is the displacement between the sites $x$ and $x^\prime$, and other non-fundamental variables are defined in Section \ref{sec:methods}. Equation \ref{eq:Gamma_def} leads to the identification of the site-projected thermal conductivity ($\zeta(x)$):

\begin{equation}
\label{eq:SPTC}
\zeta(x) = \sum_{x^\prime} \Xi(x, x^\prime).
\end{equation}

Clearly, summing $ \zeta(x) $ over all atomic sites $x$ yields the total thermal conductivity, $\kappa = \sum_{x} \zeta(x)$. This is a simple but appealing idea. It is in fact the same idea as computing the “Mulliken charge” in chemistry and a version of this was also implemented by Chetty and Martin to compute a spatially local estimate of the total energy from DFT \cite{chetty1,chetty2}. As Chetty and Martin note for their calculation, there is a “gauge invariance” to the theory: an arbitrary function can be added to Zeta, provided that the sum of that function on the sites in the cell vanishes. This implies that the SPTC is not unique. As we did for the electronic transport problem we make the simplest choice “of gauge”, and select the arbitrary function to be identically zero \cite{SPC1}. In the absence of a priori information justifying an alternate gauge, this is the only reasonable choice. We have shown in Reference \cite{SPTC} that this leads to plausible predictions for the link between structure/topology and thermal transport.

\section{The $N^2$ Method}\label{app:N2}

The $N^2$ method \cite{N2} follows directly from the random phase approximation (RPA) due to Mott \cite{MottDavis, mott1969conduction} and Hindley \cite{RPA1, RPA2}, which argue that electronic conductivity ($\sigma$) is determined by electronic activity near the Fermi energy ($\epsilon_f$). Specifically, $\sigma$ is proportional to the square of the electronic density of states ($N(E)$) around $\epsilon_f$, expressed as \cite{N2}:

\begin{equation}\label{eqn:motts_sigma} 
\sigma \propto [N(E)]^2\big|_{E \rightarrow \epsilon_f} 
\end{equation}

This approximation is valid for systems with delocalized states near $\epsilon_f$, a condition that is certainly valid in W. For a large enough supercell, the contribution of the $i^{th}$ Kohn-Sham orbital ($\epsilon_i$) to the electronic density of state at $\epsilon_f$ is:

\begin{equation}\label{eqn:N}
N(\epsilon_f) = \frac{1}{M} \sum_i \delta (\epsilon_i - \epsilon_f)
\end{equation}
where \textit{M} is the dimension of the single-particle Hamiltonian and in practice, the $\delta$ function is approximated by a Gaussian function of a selected smearing width. The spatial projection of the electronic conductivity can be expressed as \cite{N2}:

\begin{subequations}\label{eqn:N2}
\begin{gather}
\bar{\zeta}(\epsilon_f,\textbf{\textit{r}}) = \frac{1}{M^2} \sum_{i,j} \delta (\epsilon_i - \epsilon_f) \delta (\epsilon_j - \epsilon_f)|\psi_i(\textbf{\textit{r}})|^2|\psi_j(\textbf{\textit{r}})|^2\beta_{ij} \\
\beta_{ij} = \frac{1}{\int |\psi_i(\textbf{\textit{r}})|^2|\psi_j(\textbf{\textit{r}})|^2 d\textbf{\textit{r}}}.
\end{gather}
\end{subequations}

\noindent where $|\psi_i(\textbf{\textit{r}})|^2$ denotes the probability density of the $i^{th}$ Kohn-Sham orbital at the spatial grid point $\textbf{\textit{r}}$. The correction factor, $\beta_{ij}$, ensures that the volume integral of Equation \ref{eqn:N2} approximates $N^2(\epsilon_f)$ and $|\psi_i(\textbf{\textit{r}})|^2|\psi_j(\textbf{\textit{r}})|^2$ vanishes in grid regions where the wavefunctions do not overlap. Since each grid point $\textbf{\textit{r}}$ represents a localized contribution to $N^2$, their projection thus highlights electronic activity and reveals the conductive pathways within the system.

\bibliographystyle{elsarticle-num}
\bibliography{Irrad_on_W}

\end{document}